\documentclass[authoryear,preprint,review,12pt]{elsarticle}

\newcommand{\Rsun}{\mbox{$R_{\odot}$}}

\newcommand{\kms}{\mbox{km s$^{-1}$}}

\usepackage{graphics}

\journal{New Astronomy}

\begin{document}

\begin{frontmatter}



\title{Spectroscopic observations of the interacting massive binary AQ\,Cassiopea\thanks{Based on the data obtained at T\"{U}B\.{I}TAK National Observatory}}


\author[OCakirli]{C. Ibanoglu} 

\author[OCakirli]{\"{O}. \c{C}ak{\i}rl{\i}\corref{cor1}}
\ead{omur.cakirli@gmail.com,Tel:+90 (232) 3111740, Fax:+90 (232) 3731403}
\cortext[cor1]{Corresponding author}

\author[OCakirli]{E. Sipahi}

\address[OCakirli]{Ege University, Science Faculty, Astronomy and Space Sciences Dept., 35100 Bornova, \.{I}zmir, Turkey.}

\begin{abstract}
New spectroscopic observations of the double-lined eclipsing binary AQ\,Cas are presented. All available 
spectroscopic and photometric observations have been analysed for the fundamental properties of the 
components. Analyses show that the system consists of a massive primary with a mass of 
17.63$\pm$0.91 M$_{\odot}$ and radius of 13.48$\pm$0.64R$_{\odot}$ and a secondary with 
12.56$\pm$0.81 M$_{\odot}$ and radius of 23.55$\pm$0.73 R$_{\odot}$, corresponding spectral 
types of B0.5($\pm$2) II-III + B3($\pm$1) II. The secondary star fills its corresponding Roche lobe and mass 
transfer to the primary star is going on. This stream considerably does affect the photometric observations 
both starting from the second quarter up to the first contact of primary eclipse and just at the second 
maximum. Thus, the light curve is distorted and tightly depended on the wavelength of the observations. The 
available multi passband light curves have been analysed by taking the stream effects, as either hot or cool 
spots, into account. The comparison of the models and observations in the $\log~(L/L_{\odot})$ - $\log~T_{eff}$ 
and $\log~g - \log~T_{eff}$ diagrams clearly shows that the more massive star is consistent with models and 
is predicted to be close to the phase of hydrogen shell ignition. Average distance to the system is 
estimated as 4150$\pm$240 pc using the BVJHK magnitudes and V-passband extinction.
\end{abstract}

\end{frontmatter}



\section{Introduction}
{\bf A massive star with a mass higher than 8-9 M$_{\odot}$ evolves through all cycles of nuclear burning up to iron when a core 
collapse supernova is formed.}
{\bf Such high mass stars undergo a strong loss of mass by the stellar} wind which decreases their mass during nuclear evolution. 
Observations of many massive stars have shown that the mass loss is larger for the more massive stars. The mass losses are so 
large that the outer atmospheres are removed and the convective cores may become visible at the surface. In this way some 
elements processed by nuclear reactions are carried out to the surface of a star. Spectroscopic observations of massive stars at 
advanced evolutionary stages reveal that the outer layers are enhanced in helium, carbon, nitrogen and oxygen. The evolutionary 
tracks are considerably different for the mass losing stars than those at constant mass \citep{Loore92}. When a star has a close 
companion its structure and evolution are significantly changed. While mass and chemical composition of a star define its evolution in 
the case of a binary system the total mass, the mass-ratio and the orbital period would affect its evolution. The mutual interaction 
between the components may  cause to mass and angular momentum exchange which result in changes in the mass-ratio and the 
orbital period of the system. Although a number of the differences between the single and binary stars are well understood several are 
still in discussion, especially for massive binary stars.

\citet{Podsiad92} have investigated in detail how binary interaction affects the pre-supernova evolution of massive close binaries and 
the resulting supernova explosions. About 30 percent of all massive {\bf stars are substantially} affected by interactions with close binary 
companions. Further evolution of a close binary system is tightly depended upon the mass and angular momentum loss. 
\citet{Eggleton00} called attention, for the first time, that mass and subsequent angular momentum loss significantly affect binary 
evolution. \citet{Sarna93} have shown that during evolution of $\beta$\,Per 15\% of the initial total mass about two of third of 
the mass lost and 30 \% of the total angular momentum were lost. Van Rensbergen his 
collaborators \citet{vanRensbergen03} have shown that $conservative$ calculations cannot produce Algols with large mass-ratios. 

Recently \citet{Torres10} collected accurate masses and radii for the components of 95 double-lined detached binary systems. Their 
criterion was such that the mass and radius of both stars should be determined within an accuracy of 3 percent or better. The number 
of stars is decreasing in both directions towards the massive stars (M$>$10 M$_{\odot}$) and low-mass stars (M$<$1 M$_{\odot}$). 
In the last three years the number of low-mass stars for which mass and radius determined are considerably increased(see for 
example \citep{Cakirli13a,Cakirli13b}. Still there are some debates about the  formation mechanism and evolution of massive 
binaries. Therefore, we initiated a spectroscopic study for the massive binary systems.    
  
AQ\ Cas is a semi-detached, $\beta$\,Lyrae type eclipsing binary with an orbital period of 11.721 days. Its spectra were obtained by 
\citet{Struve46} and the components were classified as B3 and B9. First photometric observations made by \citet{Olson85} and 
\citet{Nha88}. Later on $UBV$ observations and $I(Kron)uvby$ light curves were published by \citet{Lee93} and \citet{Olson94}, 
respectively. There is large light variations between eclipses and light curve has a deep totality lasting about ten hours. In addition, 
light curves are significantly asymmetric which appears to tightly depended on the wavelength. About 40 years later some spectra of 
the system were obtained by \citet{Lee93}. These spectra showed the traces of both stream and rotation effects.                          
   
We report here spectroscopic observations which are added to the previous data and analysed together with the photometric 
observations for obtaining accurate masses and radii of the components. In addition, we have interpreted all available photometric 
and spectroscopic data on AQ\ Cas for understanding its present structure and evolution as well as physical properties of the stars. 
We also discuss the plausible causes of asymmetries in the observed multi passband light curves of the system. 

\section{Spectroscopic observations} 
Optical spectroscopic observations of the AQ\,Cas were obtained with the Turkish Faint Object Spectrograph Camera 
(TFOSC)\footnote{http://tug.tug.tubitak.gov.tr/rtt150\_tfosc.php} attached to the 1.5 m telescope in July, 2010 under 
good seeing conditions. Further details on the telescope and the spectrograph can be found at http://www.tug.tubitak.gov.tr. 
The wavelength coverage of each spectrum was 4000-9000 \AA~in 12 orders, with a resolving power of 
$\lambda$/$\Delta \lambda$ $\sim$7\,000 at 6563 \AA~and an average signal-to-noise ratio (S/N) was $\sim$120. We 
also obtained high S/N spectra of several early type standard stars 21\,Cyg, $\tau$\,Her, HR\,153 and 21\,Peg for use 
as templates in derivation of the radial and rotational velocities. The electronic bias was removed from each image and we used the 
{\bf 'crreject' (IRAF-immatch-imcombine-"Type of rejection") } combineoption for cosmic ray removal. Thus, the resulting spectra were largely cleaned from the cosmic rays. The echelle spectra 
were extracted and wavelength calibrated by using Fe-Ar lamp source with help of the IRAF {\sc echelle} \citep{Tonry79} package.

\subsection{Radial velocities}
To derive the radial velocities, the eleven spectra obtained for the system are cross-correlated 
against the template spectra of standard stars on an order-by-order basis using the {\sc fxcor} 
package in IRAF \citep{Simkin}. 

The spectra showed two distinct cross-correlation peaks in the quadratures, one 
for each component of the binary. Thus, both peaks are fitted independently
with a $Gaussian$ profile to measure the velocities and their errors for the individual components. If the 
two peaks appear blended, a double Gaussian was applied to the combined profile using {\it de-blend} 
function in the task. For each of the eleven observations we then determined a weighted-average radial 
velocity for each star from all orders without significant contamination by telluric absorption features. Here 
we used as weights the inverse of the variance of the radial velocity measurements in each order, as 
reported by {\sc fxcor}.

Since the orbital period of the system is considerably longer and the totality lasts about ten hours it
is hard to obtain reliable times of mid-eclipse. Therefore, only nine photoelectric times for primary eclipse
are collected from the GATEWAY data-base. A linear least-squares fit to the data yields the following ephemeris  
\[ {\rm Min\,I} = {\rm HJD}\ 2453920.897 (27) + 11^d.72107 (18) \times E \] 
where the bracketed quantity is the uncertainty in the last digits of the preceding number.

The heliocentric radial velocities for the primary (V$_p$) and the secondary (V$_s$) components are 
listed in Table\,1 , along with the dates of observations and the corresponding orbital phases computed 
with the new ephemeris given above. The velocities in this table have been corrected to 
the heliocentric reference system by adopting a radial velocity value for the template stars. The 
radial velocities are plotted against the orbital phase in Fig.\,1, together with the radial velocities
obtained by \citet{Lee93}(empty circles). The three velocities of the primary star in phases 0.12-0.16 and
nine velocities in phases 0.81-0.90 are affected from gas stream and rotation. In addition, the velocities
of both components in the orbital phase about 0.38 are systematically deviated from the orbit. Therefore,
thirteen measurements for the primary and one for the secondary are not plotted in Fig.\,1. There is no systematic
difference between our and their remaining measurements.

We analysed all the radial velocities for the initial orbital parameters using the {\sc RVSIM} 
software program \citep{kane}. Figure\,1 shows the best-fit orbital solution to the radial velocity 
data. The results of the analysis are as follow: 
$\gamma$= -16.6$\pm$2.2 \kms, $K_1$=120.9$\pm$3.9 and $K_2$=169.6$\pm$3.4 \kms with circular orbit. Using these 
values we estimate the projected orbital semi-major axis and mass ratio 
as: $a$sin$i$=67.4$\pm$1.2 \Rsun~ and $q=\frac{M_2}{M_1}$=0.713$\pm$0.027.

\begin{figure}
\includegraphics[width=8.5cm,angle=0]{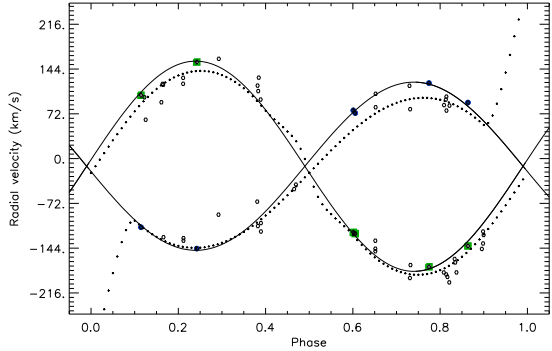}
\caption{Radial velocities for the components of AQ\,Cas. Symbols with error 
bars show the radial velocity measurements for the components of the system (primary: filled circles, secondary: open 
squares). The velocities measured by \citet{Lee93} are shown by empty circles.} \end{figure}

\begin{table}
\scriptsize
\centering
\begin{minipage}{85mm}
\caption{Heliocentric radial velocities of AQ\,Cas. The columns give the heliocentric 
Julian date, the orbital phase (according to the ephemeris in \S 2), the radial velocities of 
the two components with the corresponding standard deviations.}

\begin{tabular}{@{}ccccccccc@{}c}
\hline
HJD 2400000+ & Phase & \multicolumn{2}{c}{Star 1 }& \multicolumn{2}{c}{Star 2 } 	\\
             &       & $V_p$                      & $\sigma$                    & $V_s$   	& $\sigma$	\\
\hline
56131.4953	&	0.6010	&	77.5	&	3.2	&	-118.6	&	4.3	\\
56131.5470	&	0.6054	&	73.2	&	3.4	&	-120.6	&	4.6	\\
56133.5279	&	0.7744	&	121.4	&	2.8	&	-174.0	&	3.8	\\
56134.5740	&	0.8637	&	90.1	&	3.1	&	-140.5	&	4.2	\\
56137.5130	&	0.1144	&	-110.4	&	3.1	&	101.9	&	4.4	\\
56326.5500	&	0.2423	&	-144.7	&	2.1	&	155.3	&	3.0	\\
\hline \\
\end{tabular}
\end{minipage}
\end{table}

\subsection{Rotational velocity}
The width of the cross-correlation profile is a good tool for the measurement of $v \sin i$ (see, e.g., \citet{Queloz98}). The
rotational velocities ($v \sin i$) of the two components were obtained by measuring the FWHM of the CCF peaks in nine 
high-S/N spectra of AQ\,Cas acquired close to the quadratures, where the spectral lines have the largest Doppler-shifts. In 
order to construct a calibration curve FWHM--$v \sin i$, we have used average spectra of the $\iota$\,Psc and $\tau$\,Her. 
The FWHM of the CCF peak was measured and the FWHM-$v \sin i$ calibration was established. The $v \sin i$ values of the 
components of AQ\,Cas were derived from the FWHM of their CCF peaks. We find 287$\pm$5\,km s$^{-1}$ for the primary 
and 98$\pm$9\,km s$^{-1}$ for the secondary star.

\section{Light curves and their analyses}
The first $uvbyI$ photoelectric observations of AQ\,Cas were obtained by \citet{Olson85}. The resulting light curves are significantly 
variable in the sense that the light just before the first contact is depressed by about 0.1 mag in the u-passband relative to the fourth 
contact. This difference is decreased towards longer wavelengths. The light loss before the first contact is attributed to the stream 
seen projected against the gainer. In addition, some light fluctuations are seen beginning from mid-secondary-eclipse. Additional 
observations of AQ\,Cas were obtained by \citet{Olson89} using the same filter set. First wide passband UBV observations obtained 
between 1982 and 1989 were published by \citet{Lee93}. Despite the large fluctuations in fluxes the average shape of the light curve 
is revealed. Their BV light curves are almost symmetric except the secondary eclipse. The ingress is steeper than the egress. 
Five-color intermediate-band $uvbyI$ observations of the system were obtained by \citet{Olson94}. Thus, almost complete multi-
passband light curves were revealed. The asymmetry in the light curves was increasing towards the shorter wavelengths. In addition, 
the asymmetry is in an opposite direction as we go to the longer wavelengths. In the u-passband light curve was the most 
asymmetrical shape, being the first maximum dimmer than the second. However, the I-passband light curve is almost symmetric, the 
light level at the first maximum is slightly higher than that of the second maximum.

The observations of AQ\,Cas were also made by two familiar surveys. The R-passband light curve was obtained by the Northern Sky 
Variability Survey (NSVS) \citep{Wozniak04}. The light curve is very similar to that obtained by Olson in the $I$-passband.The last V-
passband light curve of AQ\,Cas was obtained by the  International Gamma-Ray Astrophysics Laboratory ($INTEGRAL$) mission 
\citep{Alfonso12}. With 3139 V-passband measurements the shape of the light variation is clearly defined. The magnitude of the 
system appears to reach maximum just at phase 0.75 and a steady decline up to the first contact is seen. The egress of the primary 
eclipse is steeper than that of egress. In contrary the egress of the secondary eclipse is steeper than that of egress. Both the minima 
and maxima are asymmetric. In Fig.2 we show Olson-I Lee-V, NSVS-R and $INTEGRAL$-V light curves of AQ\,Cas.     

\begin{figure}
\center
\includegraphics[width=8.5cm,angle=0]{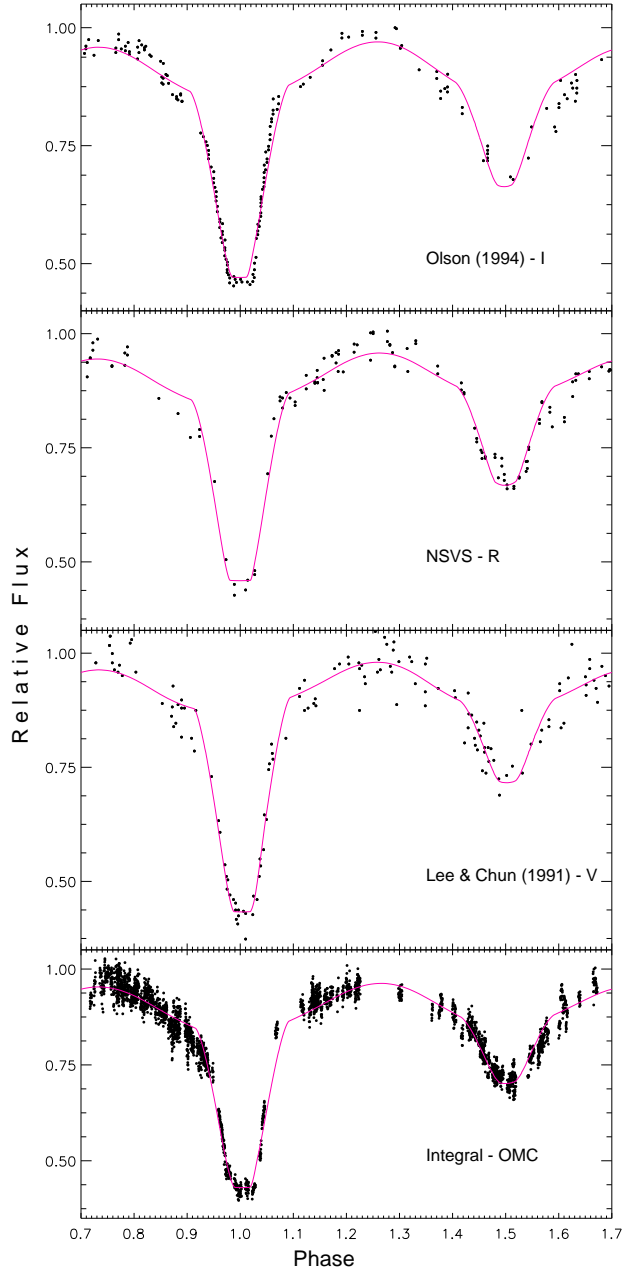}
\caption{The Olson's $I$, Lee's $V$, NSVS-$R$ and $INTEGRAL$-$V$ light curves of AQ\,Cas. The continuous lines show the best-fit model.} \end{figure}

Since the analysis of the light curves is depended on the effective temperature of the hotter component try to constrain the effective 
temperature and spectral type of the primary star using the $uvbyI$ and $JHK$ magnitudes. The intermediate passband colors of 
both components are already given by \citet{Olson85} with an interstellar reddening of E(b-y)=0.64\,mag. V=10.31, B-V=0.53, J-
H=0.182$\pm$0.039 and H-K=0.170$\pm$0.031 mag were taken from SIMBAD data-base and 2MASS catalog. Using the colors and 
reddening we estimate a spectral type and effective temperature of the primary star as B0.5III and $T_{\rm eff} = 27\,000 \pm 1000 $K 
by color-spectral type-effective temperature relations given by \citet{deJager87}, \citet{Drilling00} and \citet{Tokunaga00}.

We used the eclipsing binary light curve modeling algorithm of \citet{Wilson71}, as implemented in the {\sc phoebe} code 
of \citet{Prsa05}. The non-linear square-root limb-darkening and bolometric limb-darkening are adopted from 
\citet{Diaz95} \citet{vanham93}, respectively. The gravity-brightening coefficients $g_1$=$g_2$=1.0 and albedos $A_1$=$A_2$=1.0
were assumed for the components. The rotational velocity of the primary star is measured about 287 km s$^{-1}$, being five times
faster than synchronous rotation.  While the less massive star is appeared to be synchronous with the orbital one, for the 
more massive component $F_1$=5 is assumed. Using a trial-error method, we obtained a set of parameters which represented
the observed light curves. This preliminary analysis shows that the secondary component fills its corresponding Roche lobe.
Therefore, Mode-5 was adopted.

The shapes of the light curves are tightly depended on the wavelength. Photometric and spectroscopic observations clearly show that 
there is a mass stream from the secondary to the primary star. The stream impacts directly to the more massive star. This impact 
region is seen by the observer just at the second quarter. Therefore, additional light is observed. Thereafter, the stream begins to 
absorb the light of the primary star which causes light loss. On the other hand, during the second quarter the observer receives light 
from the steam in addition to both stars. Therefore, the brightness of the system is slightly increased.            All these effects can only 
be modelled by the hot or cool spots in the Wilson-Devinney code.  

The adjustable parameters in the light curves fitting were the orbital inclination, the surface potential 
of the primary star, the effective temperature of the secondary, and the wavelength-dependent luminosity of the hotter 
star, the zero-epoch offset. Our final results are listed in Table\,2. The uncertainties assigned to the adjusted parameters
are the internal errors provided directly by the code.  The computed light curve corresponding to the simultaneous light-velocity 
solution is compared with the observations in Fig.\,2.

\begin{table}
\scriptsize
\caption{Results of the simultaneous analyses of the Olson's $I$, Lee's $V$, NSVS-$R$ and $INTEGRAL$-$V$ light curves 
for AQ\,Cas light curves.}
\begin{tabular}{lr}
\hline
Parameters & Adopted  \\
\hline	
$i^{o}$			               			 							&84.41$\pm$0.40				\\
T$_{eff_1}$ (K)												&17\,000[Fix]						\\
T$_{eff_2}$ (K)												&16\,700$\pm$400			\\
$\Omega_1$													&5.781$\pm$0.166			\\
$\Omega_2$													&4.640$\pm$0.155		\\
$r_1$																&0.1995$\pm$0.0088		\\
$r_2$																&0.3484$\pm$0.0090		\\
$\frac{L_{1}}{(L_{1}+L_{2})}$ (Olson-$I$) 			&0.4028$\pm$0.0101			\\
$\frac{L_{1}}{(L_{1}+L_{2})}$ (Lee-$V$) 			   &0.4584$\pm$0.0114			\\
$\frac{L_{1}}{(L_{1}+L_{2})}$ ($NSVS$-R) 			&0.4190$\pm$0.0119			\\
$\frac{L_{1}}{(L_{1}+L_{2})}$ ($INTEGRAL$-V) 			&0.4424$\pm$0.0023			\\
$\sigma$															&0.032								\\				
\hline
\end{tabular}
\end{table}

Combining the results of radial velocities and light curves analyses we have calculated the absolute parameters of the stars. The 
fundamental stellar parameters for the components such as masses, radii, luminosities are listed in Table\,3 together with their formal 
standard deviations. The standard deviations of the parameters have been determined by JKTABSDIM\footnote{This can be obtained 
from http://http://www.astro.keele.ac.uk/$\sim$jkt/codes.html} code, which calculates distance and other physical parameters using 
several different sources of bolometric corrections \citep{south}. The mass for the primary of $M_P$ = 17.63 $\pm$ 0.91M $_{\odot}$  
and secondary of $M_S$ = 12.56 $\pm$ 0.81M$_{\odot}$ are consisting of a B0.5II-III star and an evolved B3 super-giant star 
\citep{Drilling00}.

\begin{table}
\scriptsize
 \setlength{\tabcolsep}{2.5pt} 
  \caption{Properties of the AQ\,Cas components}
  \label{parameters}
  \begin{tabular}{lcc}
  \hline
   Parameter 																& Primary	&	Secondary										\\   
   \hline
    Mass (M$_{\odot}$) 														& 17.63$\pm$0.91 		& 12.56$\pm$0.81			\\
   Radius (R$_{\odot}$) 														& 13.48$\pm$0.64		& 23.55$\pm$0.73			\\   
   $T_{eff}$ (K)																		& 27\,000$\pm$1000	& 16\,700$\pm$400		\\
   $\log~(L/L_{\odot})$															& 4.940$\pm$0.076		& 4.590$\pm$0.050		\\
   $\log~g$ ($cgs$) 																& 3.425$\pm$0.039 	& 2.793$\pm$0.026			\\
   Spectral Type																		& B0.5($\pm$2)II-III  	& B3($\pm$1)II    			\\
   $(vsin~i)_{obs.}$ (km s$^{-1}$)										& 287$\pm$5					& 98$\pm$9					\\ 
   $(vsin~i)_{calc.}$ (km s$^{-1}$)										& 15$\pm$1					& 4$\pm$1					\\      
   $a$ (R$_{\odot}$)																&\multicolumn{2}{c}{67.59$\pm$1.21}			\\
   $V_{\gamma}$ (km s$^{-1}$)											&\multicolumn{2}{c}{-9$\pm$1}					\\   
   $q$																						&\multicolumn{2}{c}{0.713$\pm$0.027}		\\     
   $d$ (pc)																				& \multicolumn{2}{c}{4150$\pm$240}				\\
\hline  
  \end{tabular}
\end{table}

The interstellar reddening is estimated by  \citet{Olson85} as E(b-y)=0.64 mag which corresponds to E(B-V)=0.88 mag. Using this value and the BVRJHK magnitudes   and bolometric corrections given by \citet{Girardi02} we estimated an average distance to the system as 4150$\pm$240 pc.

\section{Results and Conclusions}
First spectroscopic observations were made by \citet{Struve46} who estimated minimum mass for the components as 
11M$_{\odot}$+15M$_{\odot}$ and spectral type of B3+B9. Almost a half century later \citet{Lee93} derived absolute physical 
properties of the component stars using their spectroscopic and photometric observations. The masses and radii are found as 
18.1M$_{\odot}$+12.9M$_{\odot}$ and 13.5R$_{\odot}$+25.8R$_{\odot}$, respectively. They classify the system as a semi-detached 
eclipsing binary without any conclusion about the spectral types of the components. A year later \citet{Olson94} published the five-
color light curves and their analysis. He also measured radial velocities of the cooler star from the spectroscopic observations of OI 
$\lambda$7774 line. He estimates masses for the components 16.6M$_{\odot}$ and 4.72M$_{\odot}$ , radii of 7.55R$_{\odot}$ and 
16.6R$_{\odot}$.  \citet{Polushina04} published a catalog of massive close binaries with early-type components which includes 
AQ\,Cas. The spectral types and masses of the components are given as O8.5III+O8.5III and 29.5M$_{\odot}$+24.6M$_{\odot}$. Our 
re-analyses of all available photometric and spectroscopic data yield the mass, absolute radius and effective temperature for the 
primary and secondary stars are 17.63$\pm$0.91M$_{\odot}$, 13.48$\pm$0.64R$_{\odot}$ , 27\,000$\pm$1000, and 
12.56$\pm$0.81M$_{\odot}$, 23.55$\pm$0.73, 16\,700$\pm$400 , respectively. These values correspond to the spectral type of 
B0.5III and B3II. Thus, the masses of the components have been derived with an accuracy of 5 percent for the primary and 6 percent 
for the secondary and radii of 5 percent and 3 percent, respectively.

In Fig.3 we plot the location of the primary star of AQ\,Cas in {$\log~(L/L_{\odot})$-$\log~T_{eff}$ (left panel)and $\log~g - 
\log~T_{eff}$ (right panel) diagrams with error bars. The evolutionary tracks of a 17.63 M$_{\odot}$ have been constructed by 
\citet{Ekstrom12} models for X=0.720, Y=0.266 and Z=0.014. Stellar rotation is taken into account in their models. Despite their 
models provide a good description of the average evolution of non-interacting single stars, locations of the gainer are in good 
agreement with evolutionary tracks as seen in both panels. The primary star is predicted to be close to the phase of hydrogen shell 
ignition.

\begin{figure}
\center
\includegraphics[width=8.5cm,angle=0]{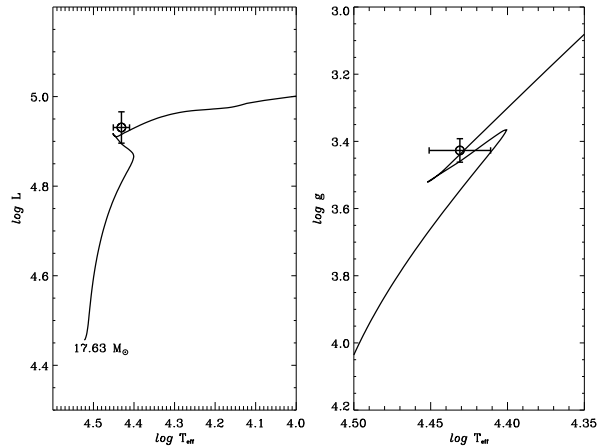}
\caption{$\log~(L/L_{\odot})$-$\log~T_{eff}$ (left panel)and $\log~g - \log~T_{eff}$ (right panel) plots of the primary component 
of AQ\,Cas compared with \citet{Ekstrom12} models for Z=0.014. } \end{figure}    

AQ\,Cas is a classical Algol. While the more massive star is filling its Roche lobe by about 0.52 the less massive star completely 
fills and overflows  its lobe. A mass transfer is taking place from the less massive (donor) star to the more massive (gainer).Mass flows 
from the inner Lagrangian point and the trajectory and properties of the mass-transferring stream are studied in detail by 
\citet{Lubow75}. The type of accretion structures can be predicted from position of a gainer in the well-known $radius-mass 
ratio$ diagram. If the orbital period of a system is too long, the fractional radius will be too small that mass transfer results in 
classical accretion disks. If the fractional radius of the gainer is sufficiently large, stream directly strikes its photosphere. This 
impact leads to the formation of a hot impact region and variable accretion structures. The gainer of the AQ\,Cas, with 
$r$=0.2 and $q$=0.7, locates in the region of impactors (see Fig.3 in \citet{Dervisoglu10}). Although its position is 
close to U\,Cep in the $r-q$ plane, the mass-ratio and orbital period for AQ\,Cas are very different. It seems that 
AQ\,Cas is a mass-transferring system with higher mass-ratio and longer orbital period. 

Recently \citet{Dervisoglu10} studied spin angular momentum evolution of the accreting components of Algol-type interacting binaries. They 
demonstrate that a small amount of mass transfer leads to accumulation of enough angular momentum which spins the gainer up to its critical 
rotation velocity. Since the gainer is in Algols have radiative envelopes, angular momentum and mass loss are not taken place due to the
enhanced magnetic activity, as are occured in solar-type chromospherically active stars. Therefore, they take into account generation of
magnetic fields in the radiative atmospheres in differentially rotating star and the possibility of angular momentum loss driven by strong stellar 
winds. Differential rotation induced by the accretion itself may produce stellar winds which carry away enough 
angular momentum to reduce their rotational velocities to the presently observed values. They suggest that this self consistent model 
should be more effective in the angular momentum loss of long-period Algols. The rotational velocity of the gainer was measured as 
240 km s$^{-1}$ by \citet{Etzel93}, and they estimate an  $F_{1}$ value of 7.3. We measured a rotational velocity of  
$v_{rot}sin~i$=287$\pm$5\,km s$^{-1}$ which is five times faster than synchronous rotation. As the mass of the gainer increases the 
rotation velocity is decreasing, as shown in their Fig.2. The $F_{1}$ value found in this study is in agreement with their estimate for 
massive primaries with orbital periods longer than 5 days. As discussed by \citet{Ibanoglu06} interacting Algols with $q>0.3$ and 
$P>5$ days have almost the same angular momentum to those of detached binaries. So, with a mass-ratio of 0.7 and an orbital 
period of 11.7 days the angular momentum loss during the evolution of AQ\,Cas is not too high. Comparison with the evolutionary tracks 
indicates similar results.

\section*{Acknowledgments}
We thank to T\"{U}B{\.I}TAK National Observatory (TUG) for a partial support in using RTT150 telescope with project 
number 11BRTT150-198.
We also thank to the staff of the Bak{\i}rl{\i}tepe observing station for their warm hospitality. This study is supported by Turkish Scientific and Technology Council under project number 112T263 and 108T237.
The following internet-based resources were used in research for this paper: the NASA Astrophysics Data System; the 
SIMBAD database operated at CDS, Strasbourg, France; T\"{U}B\.{I}TAK ULAKB{\.I}M S\"{u}reli Yay{\i}nlar 
Katalo\v{g}u-TURKEY; and the ar$\chi$iv scientific paper preprint service operated by Cornell University. 
The authors are indebted to the anonymous referee for his/her valuable suggestions which improved the paper.

\bibliographystyle{elsarticle-harv}



\end{document}